\documentclass[doublespace]{article}
\usepackage{amsmath}   
\usepackage{amsfonts}
\usepackage{amsthm}
\usepackage{amsmath, amsfonts, amssymb}
\usepackage{amsthm}
\usepackage{color}

\usepackage{ulem}

\newcommand{\calB}{\mathcal{B}}
\newtheorem{defn}{Definition}[section]
\newtheorem{theo}[defn]{Theorem}
\newtheorem{prop}[defn]{Proposition}
\newtheorem{coro}[defn]{Corollary }
\newtheorem{exam}[defn]{Example }
\theoremstyle{remark}
\newtheorem{remark}{Remark}[section]

\DeclareMathOperator{\Tr}{Tr\,}

\begin{document}

\title{Subfield-Subcodes of Generalized Toric codes}

\author{Fernando Hernando, Michael E. O'Sullivan,\\ 
Emanuel Popovici and Shraddha Srivastava\thanks{This work is supported by Science Foundation Ireland (SFI) Claude Shannon Institute, grant number 06/MI/006.
F.Hernando is also partially supported by MEC MTM2007-64704 and Junta de CyL VA025A07 (Spain).
M.~E.~O'Sullivan is  supported by the National Science
Foundation under Grant No. CCF-0916492.}}
\date{}

\maketitle

\begin{abstract}
We study subfield-subcodes of Generalized Toric (GT) codes over
$\mathbb{F}_{p^s}$. These are the multidimensional analogues of BCH
codes, which may be seen as  subfield-subcodes of generalized
Reed-Solomon codes  \cite{Cui-Pei}, \cite{delsarte},
\cite{HMS}, \cite{Tomoharu- Ryutaroh-Kohichi}, \cite{Stichtenoth}. We
identify polynomial generators for subfield-subcodes of GT codes which
allows us to determine  the dimensions and obtain bounds for the
minimum distance. We give several examples of binary and ternary
subfield-subcodes of GT codes that are the best known codes of a given
dimension and length.
\end{abstract}

\section{Generalized Toric codes}
Toric codes are algebraic geometry codes over toric varieties. These
codes were introduced by J.P. Hansen \cite{hansen2000}, see also
\cite{hansen2002}, \cite{Little-Schwarz}. Let $M$ be an integral
lattice and $P$ be a convex polytope in $M\otimes \mathbb{R}$. The
toric code $C_P$ over $\mathbb{F}_q$ associated to $P$ is the
evaluation code generated by the monomials $x^{\alpha}$ where
$\alpha\in P\cap M$ at the points of the algebraic torus
$T=(\mathbb{F}_q^*)^r$. A lower bound for the minimum distance is
estimated in \cite{diego1} using intersection theory and mixed
volumes, extending the methods of J.P. Hansen for plane polytopes.

D.Ruano introduces a natural generalization of this family,
the so called Generalized Toric Codes \cite{diego}, which consist of the evaluation of any polynomial algebra in the algebraic torus.
More precisely, one may consider any subset $U\subseteq \{0,\ldots,q-2\}^{r}$ and the corresponding
vector space $\mathbb{F}_q[U]=\langle\{x^{u}=x_1^{u_1}\cdots x_r^{u_r}\mid  u=(u_1,\ldots,u_r)\in U\}\rangle\subset \mathbb{F}_q[x_1,\ldots,x_r]$,
thus the Generalized toric code,  $C_U$, is the image under the $\mathbb{F}_q$-linear map,
$ev:\mathbb{F}_q[U] \rightarrow \mathbb{F}_q^n$, $ev(f)=(f(t))_{t\in T}$, $n=(q-1)^r$.
It is clear from his construction that any toric code is a GT code.

\begin{prop}\label{isomorphism}

Let $H=\{0,\ldots,q-2\}^r$ and $n=(q-1)^r$. The $\mathbb{F}_q$-linear map
$$
ev:\mathbb{F}_q[H]\rightarrow \mathbb{F}_q^n,
\ \ \ f\rightarrow (f(t))_{t\in T}
$$
is an isomorphism
\end{prop}
\begin{coro}
In particular, $ev$ restricted  to
$\mathbb{F}_q[U]$ is injective, so $dim(C_U)=|U|$
\end{coro}

The next result may be found in \cite{maria-michael} and \cite{diego}.
\begin{prop}\label{DualGTC}
For $u \in H$, let $\hat{u} \in H$ be defined by $\hat{u}_i= 0$ if $u_i=0$ and $\hat{u}_i=q-1-u_i$ if $u_i\ne 0$.
Let $C$ be the GT code defined by $U\subset H$, then $C^{\perp}$ is the GT code
defined by $U^{\perp}=  \{\hat{u} : u \in U\}$.

\end{prop}

\section{Subfield-Subcodes}
From now on $q=p^s$ where $p$ is a prime number.
\begin{defn}
Let $C$ be a linear code  of length $n$ over $\mathbb{F}_{p^s}$, the subfield-subcode of $C$, say $D$,
is the set of the codewords $c\in C$ such that $c\in \mathbb{F}_{p}^n$, i.e., $D=C\cap\mathbb{F}_{p}^n$.
\end{defn}
Many authors have been interested in computing the dimension of
subfield-subcodes. Delsarte studied in \cite{delsarte} the
subfield-subcodes of  modified Reed-Solomon codes. Stichtenoth
improved this lower bound in \cite{Stichtenoth} and  Shibuya et al
gave a better lower bound \cite{Tomoharu- Ryutaroh-Kohichi}. Later
on Hattori, McEliece and Solomon gave a lower bound on the dimension
of subspace-subcodes of Reed-Solomon codes. Finally Jie and Junying
generalize the previous bound for Generalized Reed-Solomon codes.

In particular Delsarte provides the following result
\cite{delsarte}:
\begin{theo}\label{delsarteTh}
$$
(C\cap\mathbb{F}_{p}^n)^{\perp}=\Tr (C^{\perp})
$$
where $\Tr :\mathbb{F}_{p^s}\rightarrow \mathbb{F}_{p}$ sending $x$ to $x+x^p+\cdots+x^{p^{s-1}}$.
\end{theo}

The next result is provided in \cite{Stichtenoth-book} although it is possibly known before.
\begin{prop}
A BCH code $D$ over $\mathbb{F}_{p}$ of length $n=p^s-1$ is a subfield-subcode of a Reed-Solomon code $C$  over
$\mathbb{F}_{p^s}$, and therefore  $d(D)\geq d(C)$.
\end{prop}

\section{Subfield subcodes of  Generalized Toric codes}
Let $R$ be $\mathbb{F}_{p^{s}}[y_1,\dots, y_r]/\langle y_1^{p^s-1}-1,\ldots,y_r^{p^s-1}-1\rangle$. We are looking for $f\in R$ such that $f(t) \in\mathbb{F}_p, \forall t\in T$. If this occurs we say that $f$ is a polynomial evaluating to $\mathbb{F}_p$. The idea is to find out first all those polynomials evaluating to $\mathbb{F}_p$ in $R$ and then restrict this set to $\mathbb{F}_{p^s}[U]$.

\begin{prop}
$ev(f)\in \mathbb{F}_p^n \Leftrightarrow f(t)=(f(t))^p ~ \forall t\in T \Leftrightarrow f^p=f$
 in $R$.
\end{prop}
\begin{proof}
According to Proposition \ref{isomorphism} we know that $ev(f)^p=ev(f^p)$ then it is clear that:

$ev(f)\in \mathbb{F}_p^n \Leftrightarrow f(t)=(f(t))^p  \forall t\in T \Leftrightarrow
ev(f)=ev(f)^p \Leftrightarrow ev(f)=ev(f^p) \Leftrightarrow ev(f-f^p)=0 \Leftrightarrow
f-f^p\in ker(ev) \Leftrightarrow f^p(\underline{y})=f(\underline{y})$ in $R$.
\end{proof}

Consider $G=Gal(\mathbb{F}_{p^s}  \mid \mathbb{F}_{p})=\{g_0,\ldots,g_{s-1}\}$  the Galois group, where
$g_i$ maps $\alpha$ to $\alpha^{p^i}$. Looking at exponents, we may consider $G$ to act on $\mathbb{Z}_{p^s-1}$ by multiplying by $p$ and this may be naturally extended to $\mathbb{Z}_{p^s-1}\times\cdots\times \mathbb{Z}_{p^s-1}$ by multiplying by $p$ coordinate wise. The orbits
of $G$ on $\mathbb{Z}_{p^s-1}\times\cdots\times \mathbb{Z}_{p^s-1}$ are called cyclotomic cosets, i.e., for every $\underline{b}\in \left( \mathbb{Z}_{p^s-1}\right)^r$
we define the cyclotomic coset $I_{\underline{b}}$ by $\{\underline{b},p\underline{b},\ldots,p^{n_{\underline{b}}-1} \underline{b}\}$ where $n_{\underline{b}}$ is the smallest positive integer such that $\underline{b}= \underline{b}p^{n_{\underline{b}}} $. The integer $n_{\underline{b}}$ is the cardinal of $I_{\underline{b}}$.

Some known  properties of cyclotomic cosets:

\begin{prop}\label{propertiescoset} \ \ \ \ \ \ \ \ \ \ \
\begin{itemize}
\item[(i)] $I_{\underline{b}}=\{\underline{b},p\underline{b},p^2\underline{b},\ldots,p^{n_{\underline{b}}-1}\underline{b}\}$ is closed under multiplication by $p$.
\item[(ii)] The cardinal of $I_{\underline{b}}$ is either $s$ or a divisor of it.
\item[(iii)] $I_{\underline{b}}$ and $ I_{\underline{b}'}$ are either identical or they don't intersect.
Thus  $\calB =  \{ I_{\underline{b}} :{\underline{b}} \in \left( \mathbb{Z}_{p^s-1}\right)^r\}$ partitions
$\left( \mathbb{Z}_{p^s-1}\right)^r$.
\end{itemize}
\end{prop}

If $\theta:R \rightarrow R$ is an isomorphism and $f$ evaluates to $\mathbb{F}_p$. Then so does $\theta(f)$. This is because
$\theta(f)^p=\theta(f^p)=\theta(f)$. So it is worthwhile cataloguing some isomorphisms of $R$.

\begin{prop}\label{IsomorphismR}
\begin{itemize}
\item[(i)] For any $i$ coprime with $p^{s}-1$, the map $\theta_i$ fixing $F_{p^s}$ and taking $f(y_1,\ldots,y_r)\rightarrow f(y_1^i,\ldots,y_r^i)$ is an isomorphism of $R$.

\item[(ii)] For any $\underline{\alpha}\in\mathbb{F}_{p^s}^*\times\cdots\times\mathbb{F}_{p^s}^*$, the map $\theta_{\underline{\alpha}}$ fixing $\mathbb{F}_{p^s}$ and taking $f(y_1,\ldots,y_r)\rightarrow f(\alpha_1 y_1,\ldots,\alpha_ry_r)$ is an
isomorphism of $R$.

\item[(iii)] The Frobenious map on $\mathbb{F}_{p^s}$ combined with $y_i\mapsto y_i$ for $i =1, \dots,r$.
\end{itemize}
\end{prop}

Let $f(\underline{y})=\sum a_{i_1,\ldots,i_r} y_1^{i_1}\cdots y_r^{i_r} \in R$, we denote $supp(f)=\{\underline{i}\mid a_{\underline{i}}\neq 0\}$ as the support of $f$. If $I_{\underline{b}}$
is a cyclotomic coset, we denote $f_{I_{\underline{b}}}=\sum_{\underline{i}\in I_{\underline{b}}} \underline{y}^{\underline{i}}$ as the polynomial having $supp(f)=I_{\underline{b}}$ and coefficients equal to one.

It is easy to verify that $f_{I_{\underline{b}}}$ evaluates to
$\mathbb{F}_p$.
Note that
$\theta_{\underline{\alpha}}(f_{I_{\underline{b}}})=\sum_{\underline{i}\in   I_{\underline{b}}} \alpha_1^{i_1} y_1^{i_1}\cdots \alpha_r^{i_r}
y_r^{i_r}$
is the polynomial with support $I_{\underline{b}}$ and coefficients determined by $\underline{\alpha}$.
Since  $\theta_{\underline{\alpha}}$ is an isomorphism,  $\theta_{\underline{\alpha}}(f_{I_{\underline{b}}})$
evaluates to $\mathbb{F}_p$.

Let $l= |\calB|$ be the number of cyclotomic cosets and let
$J=\{\underline{b}_1,\ldots,\underline{b}_l\}$, be a set of representatives, so
$\calB=\{I_{\underline{b}_1},\ldots,I_{\underline{b}_l}\}$.
From now on we will denote by $f_{I_{\underline{b}},\beta}$  the polynomial with support $I_b$ and
leading coefficient $\beta$, i.e.,
$f_{I_{\underline{b}},\beta}=\beta \underline{y}^{\underline{b}}+\beta^p\underline{y}^{\underline{b}p}+\cdots+\beta^{p^{n_{\underline{b}-1}}}\underline{y}^{\underline{b}p^{n_{\underline{b}-1}}}$.
Note that $f_{I_{\underline{b}}, \beta}$ evaluates to
$\mathbb{F}_p$ if and only if  $\beta\in \mathbb{F}_{p^{n_{\underline{b}}}}$.

\begin{prop}\label{LI}
Let  $f$ be a function that evaluates to $\mathbb{F}_p$ with $supp(f)=I_{\underline{b}}$ and let $\beta$ be a primitive element of $\mathbb{F}_{p^{n_{\underline{b}}}}$. Then, $f$ is a linear combination of $f_{I_{\underline{b}}},f_{I_{\underline{b}},\beta},\ldots,f_{I_{\underline{b}},\beta^{n_b-1}}$.
\end{prop}
\begin{proof}
Since $supp(f)=I_{\underline{b}}$ and $f^p=f$ there is some $\alpha$ such that  $f=\alpha \underline{y}^{\underline{b}}+\alpha^{p}\underline{y}^{\underline{b}p}+\cdots+\alpha^{p^{n_{\underline{b}-1}}}\underline{y}^{\underline{b}p^{n_{\underline{b}-1}}}$.
Moreover $ \alpha^{p^{n_{\underline{b}}}}  = \alpha$, which implies that $\alpha \in \mathbb{F}_{p^{n_{\underline{b}}}}$.

We know that $\{1,\beta, \ldots,\beta^{(n_{\underline{b}}-1)}\}$ is a basis of
$\mathbb{F}_{p^{n_{\underline{b}}}}$ over  $\mathbb{F}_{p}$, so
$\alpha=a_0+a_1\beta+\cdots+a_{n_{\underline{b}}-1}\beta^{(n_{\underline{b}}-1)}$,
with $a_i\in \mathbb{F}_p$ for all $i$. Therefore,

\begin{align*}
 f &= \sum_{i=0}^{n_{\underline{b}}-1} \alpha^{p^i} \underline{y}^{\underline{b}p^i}
 = \sum_{i=0}^{n_{\underline{b}}-1}  \underline{y}^{\underline{b}p^i}
\Big( \sum_{j=0}^{n_{\underline{b}}-1} a_j\beta^j\Big)^{p^i}\\
&= \sum_{j=0}^{n_{\underline{b}}-1} a_j
\sum_{i=0}^{n_{\underline{b}}-1} \beta^{jp^i} \underline{y}^{\underline{b}p^i} 
= \sum_{j=0}^{n_{\underline{b}}-1} a_j f_{I_{\underline{b}}, \beta^j}
\end{align*}
\end{proof}

\begin{prop}\label{LinearlyInd}
$f_{I_{\underline{b}}},f_{I_{\underline{b}},\beta},
\ldots,f_{I_{\underline{b}},\beta^{n_{\underline{b}-1}}}$ are linearly independent over $\mathbb{F}_p$.
\end{prop}
\begin{proof}
Suppose it is not true. Thus, $a_0f_{I_{\underline{b}}}+a_1f_{I_{\underline{b}},\beta}+\cdots+a_{n_{\underline{b}}-1}f_{I_{\underline{b}},
{\beta^{n_{\underline{b}}-1}}}=0$. The smallest monomial  in the left hand side is
 $(a_0+a_1\beta+\cdots+a_{n_b-1}\beta^{(n_{\underline{b}}-1)})\underline{y}^{\underline{b}}$ which has to be zero. This is true if $\beta$ is a root of $p(z)=a_0+a_1z+\cdots+a_{n_b-1}z^{n_b-1}$, but this is not possible because the minimal polynomial of $\beta$ has degree $n_{\underline{b}}$ .
\end{proof}

\begin{theo}\label{base}

A basis for the set of polynomials evaluating to $\mathbb{F}_p$ is:

\[\bigcup_{I_{\underline{b}}\in \calB} \{f_{I_{\underline{b}},\beta^j}: j \in \{0,\dots,n_{\underline{b}}-1\}, \beta \text{ primitive in }\mathbb{F}_{p^{n_{\underline{b}}}}\}.\]
\end{theo}
\begin{proof}
If $I_{\underline{b}}$ and $I_{\underline{b}'}$ are two different cosets then $f_{I_{\underline{b}},\beta}$ and  $f_{I_{\underline{b}'},\beta'}$
have different degrees. So, there is no way to generate one from the other which proves that different classes are linearly independent.
Moreover within the set of polynomials with the same support, say $I_b$, we know from Corollary \ref{LinearlyInd} that the only linearly independent are $\{f_{I_b,1},f_{I_b,{\beta}},\ldots,f_{I_b,{\beta^{n_b-1}}}\}$. So, the only part left is to see that it is a system of generators.

Consider the smallest monomial in $f$, say $\beta^{j_1} \underline{y}^{\underline{b}}$ then $f_{I_{\underline{b}},\beta^{j_1}}=\sum_{k=0}^{n_{b}-1}(\beta^{j_1}\underline{y}^{\underline{b}})^{p^k}$
must appear in $f$. Since $\beta^{j_1} \underline{y}^{\underline{b}}$ is the smallest monomial in $f$, therefore ${\underline{b}}$ must be one of the leaders in $J=\{\underline{b}_1,\ldots,\underline{b}_l\}$. Assume without
loss of generality that $\underline{b}_1<\underline{b}_2<\cdots<\underline{b}_l$ and  $\underline{b}=\underline{b}_1$.

Consider $f_1=f-f_{I_{\underline{b}_1},\beta^{j_1}}$ and the first monomial on it, say $\beta^{j_2}\underline{y}^{\underline{b}'}$. Again, the polynomial
$f_{I_{\underline{b}'},\beta^{j_2}}=\sum_{k=0}^{n_{b}-1}(\beta^{j_2}\underline{y}^{\underline{b}'})^{p^k}$ must appear in $f_1$. We may assume that $\underline{b}'=b_2$ and consider
$f_2=f_1-f_{I_{\underline{b}_2},\beta^{j_2}}$.

In at most $l$-steps, we can finish the process obtaining that $f=a_1f_{I_{\underline{b}_1},\beta^{j_1}}+\cdots+a_lf_{I_{\underline{b}_l},\beta^{j_l}}$,
which concludes the proof.
\end{proof}

For the next result we introduce  an $\mathbb{F}_p$ linear mapping on $R$ extending the trace map,
 $T: R \rightarrow R$ is given by $g \mapsto g+ g^p + \dots
 g^{p^{s-1}}$ for all $g \in R$.
\begin{coro}
The image of $T$ is exactly the set of $f\in R$ that evaluate to $\mathbb{F}_p$.
\end{coro}
\begin{proof}
Let  $f=T(g)=g+g^p+\cdots+g^{p^{s-1}}$.  Since $g^{p^s}=g$ in $R$ we have
$f^p = f$. Thus any image of the map $T$ evaluates to $\mathbb{F}_p$.

For the converse, it is sufficient, by Proposition~\ref{LI}, to show
that each $f_{I_{\underline{b}_l},\beta}$  is in the image of $T$ for
$\beta$ an element of $\mathbb{F}_{p^{n_{\underline{b}}}}$.  Let
$\gamma\in \mathbb{F}_{p^s}$ be such that
$\Tr _{\mathbb{F}_{p^s}   /\mathbb{F}_{p^{n_{\underline{b}}}}}(\gamma)
= \beta$.  Then
\begin{align*}
T(\gamma \underline{y}^{\underline{b}})
&=\sum_{i=0}^{s-1} \gamma^{p^i}  \underline{y}^{\underline{b} p^i} \\
&= \sum_{j=0}^{\frac{s}{n_{\underline{b}}}-1}
\sum_{i=0}^{n_{\underline{b}}-1} \gamma^{p^{i+jn_{\underline{b}}}}
 \underline{y}^{\underline{b}p^{i+jn_{\underline{b}}}}\\
\intertext{Since $\underline{b}p^{n_{\underline{b}}} =
  \underline{b}$, }
&= \sum_{i=0}^{n_{\underline{b}}-1} \underline{y}^{\underline{b}p^{i}}
\Big(\sum_{j=0}^{\frac{s}{n_{\underline{b}}}-1} \gamma^{(p^{n_{\underline{b}}})^j} \Big)^{p^i}
\intertext{The term in parentheses is
 $\Tr _{\mathbb{F}_{p^s}   /\mathbb{F}_{p^{n_{\underline{b}}}}}(\gamma)
 = \beta$, so }
T(\gamma \underline{y}^{\underline{b}})  &=
f_{I_{\underline{b}_l},\beta}
\end{align*}
\end{proof}

This provides us a constructive way of producing all those polynomials
which evaluate to $\mathbb{F}_{p}$. In particular, if we restrict to
those polynomials with support in $U$, we trivially have a formula for
the dimension of a subfield-subcode.
\begin{theo}
\label{t:D_U}
Let $U \subseteq \{0\dots,q-2\}^r$ and let $D_U = C_U \cap \mathbb{F}_p^n$.
\[D_U = ev\Big(T(\mathbb{F}_{p^s}[H])\cap\mathbb{F}_{p^s}[U]\Big)
\]
A basis for $D_U$ is:
\[\bigcup_{I_{\underline{b}}:I_{\underline{b}}\subseteq U} \{f_{I_{\underline{b}},\beta^j}: j\in\{0,\ldots,n_{\underline{b}}-1\}
, \beta \text{ primitive in }
\mathbb{F}_{p^{n_{\underline{b}}}}\}\]
Moreover it has dimension
 \[ \dim D_U=\sum_{I_{\underline{b}}:I_{\underline{b}}\subseteq U} n_{{\underline{b}}} \].

\end{theo}

\begin{remark}

When $r=1$ and  $U=\{0,1,2,\ldots,k-1\}$ the GT code is a Reed-Solomon code with parameters $[p^s-1,k,p^s-k]$.
\end{remark}

\begin{exam}
Let $C$ be an $[n,k,d]$ Reed-Solomon code with $q=2^4$, $n=15$, i.e. we evaluate all the polynomials of degree less than or equal to $k-1$, at all the points of $\mathbb{F}_{16}^{\ast}$. Let $D$ be the subfield-subcode of $C$, that is, $D=C\cap\mathbb{F}_{2}^{15}$.
What are the functions $f:\mathbb{F}_{16}\rightarrow \mathbb{F}_{2}$ we have to evaluate to get $D$?

The different cosets are $I_{0}=\{0\}$, $I_1=\{1,2,4,8\}$, $I_3=\{3,6,12,9\}$, $I_5=\{5,10\}$, $I_7=\{7,14,13,11\}$. Depending on the value of $k$ we have:
\begin{itemize}
\item From $1\leq k\leq 8$, the only function is $f=1$ corresponding to the coset $I_0$, so the code $D$ is $[15,1,15]$.

\item If $k=9$, $C=[15,9,7]$ then we have $T_r(x)=f_{I_1},f_{I_1,\alpha},f_{I_1,\alpha^2},f_{I_1,\alpha^3}$ and $f_{I_0}=1$. Then $D$ is a $[15,5,7]$ code.
\item If $k=10$ nothing new.

\item If $k=11$, $C=[15,11,5]$ we consider $I_0$, $I_1$ and $I_5$. That is
$f_{I_5}=x^5+x^{10}$ and $f_{I_1,\alpha}=\alpha^5 x^5+\alpha^{10} x^{10}$ in addition to the previous functions. Therefore, $D=[15,7,5]$.

\item If $k=12$ nothing new.

\item If $k=13$, $C=[15,13,3]$ we  consider $I_0$, $I_1$, $I_5$ and $I_3$. That is
$f_{I_3,\alpha^i}=\alpha^{3i}x^3+\alpha^{6i}x^6+\alpha^{9i}x^9+\alpha^{12i}x^{12}$ in addition to the previous functions, for $0\leq i\leq 3$.
Therefore, $D=[15,11,3]$.
\item If $k=14$ nothing new.

\item If $k=15$, $C=[15,15,1]$ and $D=[15,15,1]$ with the $4$ new functions corresponding to $I_7$:
$f_{I_7,\alpha^i}=\alpha^{7i}x^7+\alpha^{11i}x^{11}+\alpha^{13i}x^{13}+\alpha^{14i}x^{14}$ for $0\leq i\leq 3$.
\end{itemize}
\end{exam}

\section{Dual of Subfield-Subcodes}

Theorem \ref{delsarteTh} together with Theorem~\ref{t:D_U} motivate this section.

Let $U \subseteq \{0,\dots,q-2\}^r$ and let $C_U= ev(\mathbb{F}_{p^s}[U])$ and $D_U = C_U \cap \mathbb{F}_p^n$.
From Proposition \ref{DualGTC}, we know that $C_U^{\perp}$ is the GT code defined by $U^{\perp}$. From  Delsarte's Theorem we have
\[ D_U^{\perp}=\Tr (C_{U^{\perp}}) = \Tr (ev (\mathbb{F}_{p^s}[U^\perp])) = ev (T(\mathbb{F}_{p^s}[U^{\perp}]))\]
The last equality follows from  $ev\circ T = \Tr \circ ev$, which is easily verified.
Clearly,  $T( \mathbb{F}_{p^s}[U^{\perp}])$ is spanned by $T(\gamma y^{\underline{b}})$ for $\underline {b}\in U^{\perp}$ and $\gamma \in \mathbb{F}_{p^s}$.  For $\underline {b}$ fixed and varying $\gamma$ we get exactly the set of  $f_{I_{\underline{b}}, \beta}$ for $\beta \in \mathbb{F}_{p^{n_{\underline{b}}}}$.
Thus we have a basis for $D_U^{\perp}$.

\begin{theo}\label{DualOfSFSC}
 $D_U^{\perp}$ has the basis
 \[ \bigcup_{I_{\underline{b}}: I_{\underline{b}}\cap U^\perp\ne \emptyset}
 \{f_{I_{\underline{b}},\beta^j}: j \in \{0,\dots,n_{\underline{b}}-1\}, \beta \text{ primitive in }\mathbb{F}_{p^{n_{\underline{b}}}}\}
 \]

 We therefore have

 \[ \dim D_U^{\perp} = \sum _{I_{\underline{b}}: I_{\underline{b}}\cap U^\perp\ne \emptyset} n_{{\underline{b}}}.\]
\end{theo}

\begin{prop}
Let $\hat{U}=\{ supp(h)\mid h=\Tr (\underline{y}^{\underline{b}}),\underline{b}\in U^{\perp} \}=
\{ I_{\underline{b}}\mid \underline{b}\in U^{\perp} \}=\{p^i\underline{b}\mid \underline{b}\in U^{\perp}, i=0,1,\ldots,n_{\underline{b}}-1\}$
Then $D_U^{\perp} =  C_{\hat{U}}\cap \mathbb{F}_p^n=D_{\hat{U}}$.
\end{prop}

\begin{coro}
One can always decode $D^{\perp}$ up to $t=\lfloor d(C_{\hat{U}})-1/2 \rfloor$ with the decoding algorithm for $C_{\hat{U}}$.
\end{coro}

\section{Computations}
From the practical point of view it makes  sense to choose $U$ to be the union of different cyclotomic cosets,
otherwise the evaluation will not be in $\mathbb{F}_{p}^n$.

We have written a Magma function for computing the subfield-subcode of a GT code and we have found
a number of optimal codes. Consider first the field $GF(2^3)$ and $r=2$ so $T$ is the toric surface.
In each of the following cases we give  a subset $U$ of $\left(\mathbb{Z}_7\right)^2$ and the parameters of $D=D_U$ and $D^\perp= D_U^\perp$, the subfield-subcode of  $C_U$ and its dual.

\vspace{.3cm}
\begin{itemize}
\item[i)] $U=[[ 1, 0 ],[ 2, 0 ],[ 4, 0 ],[ 0, 1 ],[ 0, 2 ], [ 0, 4 ]].$ \\
$D$ is $[49,6,24]$ and $D^{\perp}$ is $[49,43,3]$.

\item[ii)] $   U=[[ 6, 3 ],[ 5, 6 ],[ 3, 5 ] , [ 3, 1 ],[ 6, 2 ],[ 5, 4 ] , [ 6, 1 ],[ 5, 2 ],[ 3, 4 ]].$\\
$D$ is $[49,9,20]$ and $D^{\perp}$ is $[49,39,3]$.

\item[iii)] $U=[ [2, 1 ],[ 4, 2 ],[ 1, 4 ], [ 3, 1 ],[ 6, 2 ],[ 5, 4 ],[ 4, 1 ],[ 1, 2 ],[ 2, 4 ],[0,0]].$ \\
$D$ is  $[49,10,20]$ and $D^{\perp}$ is $[49,39,4]$.
If we consider $U'=U\cup\{[[1,0],[2,0],$\\ $[ 5,0],[6,0],[1,1],[2,2]\}$ we get a new toric code, $C_{U'}$, with parameters $[49,16,18]$, i.e, the minimum distance drops by $2$ (with respect to $C_U$) and the subfield-subcode $D_{U'}$ is equal to $D_U$. The previous is an example of a subfield-subcode $D_{U'}$ of a GT code $C_{U'}$ where $d(D_{U'})>d(C_{U'})$.

\item[iv)] $U=[[ 1, 0 ],[ 2, 0 ],[ 4, 0 ] , [ 2, 3 ],[ 4, 6 ],[ 1, 5 ], [ 0, 1 ],[ 0, 2 ],[ 0, 4 ] , [ 6, 3 ],  [ 5, 6 ],[ 3, 5 ]\\ , [ 6, 1 ],[ 5, 2 ], [ 3, 4 ]].$\\
$D$ is $[49,15,16]$ and $D^{\perp}$ is $[49,34,6]$.

\item[v)] $U=[[ 1, 0 ],[ 2, 0 ],[ 4, 0 ],[ 0, 1 ],[ 0, 2 ],[ 0, 4 ],[ 1, 1 ],[ 2, 2 ],[ 4, 4 ],[ 2, 1 ],
[ 4, 2 ],[ 1, 4 ]\\,[ 3, 1 ],[ 6, 2 ],[ 5, 4 ],[ 4, 1 ],[ 1, 2 ]\\,[ 2, 4 ],[ 1, 3 ],[ 2, 6 ],[ 4, 5 ]].$ \\
$D$ is $[49,21,12]$ and $D^{\perp}$ is $[49,28,7]$.
We use again the same strategy of adding points: consider  $U'=U\cup \{[3,0],[6,0],[6,1],[5,2]\}$, we obtain the GT code $C_{U'}$ with parameters $[49,25,9]$ where the minimum distance drops by $3$ and the subfield-subcode $D_U=D_{U'}$.

\item[vi)] $  U=[[ 6, 3 ],[ 5, 6 ],[ 3, 5 ] , [ 1, 0 ],[ 2, 0 ],[ 4, 0 ] , [ 3, 0 ], [ 6, 0 ], [ 5, 0 ] ,[ 2, 1 ],[ 4, 2 ],[ 1, 4 ]\\ , [ 3, 1 ],[ 6, 2 ],[ 5, 4 ] , [ 4, 1 ], [ 1, 2 ], [ 2, 4 ] , [ 5, 1 ],[ 3, 2 ],[ 6, 4 ] , [ 1, 3 ],[ 2, 6 ], [ 4, 5 ] , [ 2, 3 ]\\,[ 4, 6 ], [ 1, 5 ] , [ 3, 3 ], [ 6, 6 ],[ 5, 5 ] , [ 4, 3 ],[ 1, 6 ],[ 2,5]].$\\
$D$ is $[49,33,6]$ and $D^{\perp}$ is $[49,16,7]$.

\item[vii)] $ U= [[ 6, 3 ],[ 5, 6 ],[ 3, 5 ] , [ 0, 0 ] , [ 0, 1 ],[ 0, 2 ],[ 0, 4 ] , [ 1, 1 ],[ 2, 2 ],[ 4, 4 ] , [ 3, 1 ],[ 6, 2 ]\\,[ 5, 4 ] , [5, 1 ],[ 3, 2 ],[ 6, 4 ] , [ 6, 1 ],[ 5, 2 ],[ 3, 4 ], [ 0, 3 ],[ 0, 6 ],[ 0, 5 ] , [ 2, 3 ],[ 4, 6 ],[ 1, 5 ]\\ , [ 3, 3 ],[ 6, 6 ],[ 5, 5 ] , [ 4, 3 ],[ 1, 6 ],[ 2, 5 ] , [5, 3 ],[ 3, 6 ],[ 6, 5 ]].$\\
$D$ is $[49,34,6]$ and $D^{\perp}$ is $[49, 15, 12]$.

\item[viii)]	  $U=[[ 6, 3 ],[ 5, 6 ],[ 3, 5 ] , [ 0, 0 ] , [ 1, 0 ],[ 2, 0 ],[ 4, 0 ] , [ 3, 0 ], [ 6, 0 ],[ 5, 0 ] , [ 1, 1 ],[ 2, 2 ]\\,[ 4, 4 ] , [ 2, 1 ],[ 4, 2 ],[ 1, 4 ] , [ 4, 1 ],[ 1, 2 ],[ 2, 4 ] , [ 5, 1 ],[ 3, 2 ],[ 6, 4 ] , [ 6, 1 ],[ 5, 2 ],[ 3, 4 ]\\ , [ 0, 3 ],[ 0, 6 ],[ 0, 5 ] ,[ 1, 3 ],[ 2, 6 ],[ 4, 5 ] , [ 3, 3 ],[ 6, 6 ],[ 5, 5 ] , [ 4, 3 ],[ 1, 6 ],[ 2, 5 ] , [ 5, 3 ]\\,[ 3, 6 ],[ 6, 5 ]].$\\
$D$ is $[49,40,4]$ and $D^{\perp}$ is $[49, 9, 14]$.

\item[ix)]   $U=[[ 0, 0 ],[ 1, 0 ],[ 2, 0 ],[ 4, 0 ],[ 3, 0 ],[ 6, 0 ],[ 5, 0 ],[ 0, 1 ],[ 0, 2 ],[ 0, 4 ],
[ 1, 1 ],[ 2, 2 ]\\,[ 4, 4 ],[ 2, 1 ],[ 4, 2 ],[ 1, 4 ],[ 3, 1 ],[ 6, 2 ],[ 5, 4 ],[ 4, 1 ],[ 1, 2 ],
[ 2, 4 ],[ 5, 1 ],[ 3, 2 ],[ 6, 4 ]\\,[ 6, 1 ],[ 5, 2 ],[ 3, 4 ],[ 0, 3 ],[ 0, 6 ],[ 0, 5 ],[ 1, 3 ],
[ 2, 6 ],[ 4, 5 ],[ 2, 3 ],[ 4, 6 ],[ 1, 5 ],[ 3, 3 ]\\,[ 6, 6 ],[ 5, 5 ],[ 4, 3 ],[ 1, 6 ],[ 2, 5 ],
[ 5, 3 ],[ 3, 6 ],[ 6, 5 ]].$\\
$D$ is $[49,46,2]$ and $D^{\perp}$ is $[49,3,28]$.

\end{itemize}

Notice that $p=2\nmid s=3$ thus from Theorem \ref{DualOfSFSC} we know that the dual of a subfield-subcode is again the subfield-subcode of another toric code. In each example the code $D$  is the best known code for a fixed length and dimension. Also in each example, except  vi),vii) and viii) the dual code has the same correction capability as  the best known code for a fixed length and dimension.

From now on we will denote by $D$ the subfield-subcode of the GT codes over $GF(3^2)$  and $r=2$. In each of the following cases we give  a subset $U$ of $\left(\mathbb{Z}_8\right)^2$ and the parameters of $D=D_U$ and $D^\perp= D_U^\perp$, the subfield-subcode of  $C_U$ and its dual.
\vspace{.3cm}
\begin{itemize}
\item[i)] $ U=[[ 5, 0 ],[ 7, 0 ] , [ 5, 5 ],[ 7, 7 ]]$\\
         $D$ is $[64,4,42]$ and $D^{\perp}$ is $[64, 60, 2]$.
\item[ii)] $U=[[ 5, 1 ],[ 7, 3 ] , [ 0, 0 ],[ 0, 0 ] , [ 7, 1 ],[ 5, 3 ] , [ 1, 2 ],[ 3, 6] , [ 2, 1 ],[ 6, 3 ]]$\\
          $D$ is $[64,9,36]$ and $D^{\perp}$ is $[64, 55, 4]$.
\item[iii)] $U=[[ 7, 1 ],[ 5, 3 ] , [ 5, 0 ],[ 7, 0 ] , [  0, 1 ],[ 0, 3 ] , [ 1, 5 ],[ 3, 7 ] , [ 2, 1 ],[ 6, 3 ] , [ 6, 2 ],[ 2, 6 ]].$\\
          $D$ is $[64,12,30]$ and $D^{\perp}$ is $[64, 52, 4]$.

\item[iv)] $U=[[0,0],[4,0],[0,4],[4,4],[ 5, 0 ],[ 7, 0 ],[ 0, 1 ],[ 0, 3 ],[ 1, 1 ],[ 3, 3 ],[ 2, 1 ],[ 6, 3 ],
[ 3, 1]\\,[ 1, 3 ],[ 4, 1 ],[ 4, 3 ],[ 5, 1 ],[ 7, 3 ],[ 6, 1 ],[ 2, 3 ],[ 1, 2 ],[ 3, 6 ],[ 2, 2 ],[ 6, 6 ],
[ 3, 2 ],[ 1, 6 ],[ 4, 2]\\,[ 4, 6 ], [ 5, 2 ],[ 7, 6 ],[ 6, 2 ],[ 2, 6 ],[ 7, 2 ],[ 5, 6 ],[ 1, 4 ],[ 3, 4 ],
[ 2, 4 ],[ 6, 4 ],[ 0, 5 ],[ 0, 7 ],[ 5, 4 ]\\,[ 7, 4 ],[ 1, 5 ],[ 3, 7 ],[ 2, 5 ],[ 6, 7 ],[ 3, 5 ],[ 1, 7 ],
[ 7, 5 ]\\,[ 5, 7 ]]$ \\
$D$ is $[64,50,5]$ and $D^{\perp}$is $[64, 14, 27]$.
Consider $U'=U\cup \{[1,0],[6,0],[6,5],[7,7],[4,7]\}$ the new GT code $C_{U'}$ has parameters $[64, 55, 4]$ where the minimum distance drops by $1$ but $D_U=D_{U'}$.
\end{itemize}
In all the examples the code $D$  has the same correction capability to the best known codes for a fixed length and dimension.

{}

\end{document}